\documentclass[a4paper, superscriptaddress, aps, prl, reprint]{revtex4-2}

\usepackage{amsfonts}
\usepackage{amssymb}
\usepackage{amsmath}
\usepackage[utf8]{inputenc}
\usepackage[T1]{fontenc}
\usepackage[english]{babel} 
\usepackage{graphicx}
\usepackage{hyperref}
\hypersetup{colorlinks, citecolor=black, linkcolor=black, urlcolor=black}
\usepackage{comment}
\usepackage{siunitx}
\usepackage{float}
\usepackage{placeins}
\usepackage{subcaption}
\usepackage[font=small,labelfont=bf]{caption}
\usepackage{textcomp, gensymb} 
\usepackage[version=3]{mhchem}
\usepackage{multirow}
 
\begin{document}
\title{ Neutron phase filtering for separating phase- and attenuation signal in aluminium and anodic aluminium oxide } 
\author{Estrid Buhl Naver}
\affiliation{ Department of Energy Conversion and Storage, Technical University of Denmark, Kgs. Lyngby, Denmark}

\author{Okan Yetik} 
\affiliation{Laboratory for Neutron Scattering and Imaging, Paul Scherrer Institut, Villigen, Switzerland} 

\author{Noémie Ott}
\affiliation{Institute for Microtechnology and Photonics, OST - East Switzerland University for Applied Sciences, Buchs, Switzerland}

\author{Matteo Busi}
\affiliation{Laboratory for Neutron Scattering and Imaging, Paul Scherrer Institut, Villigen, Switzerland}

\author{Pavel Trtik}
\affiliation{Laboratory for Neutron Scattering and Imaging, Paul Scherrer Institut, Villigen, Switzerland}

\author{Luise Theil Kuhn}
\affiliation{ Department of Energy Conversion and Storage, Technical University of Denmark, Kgs. Lyngby, Denmark}

\author{Markus Strobl}
\affiliation{Laboratory for Neutron Scattering and Imaging, Paul Scherrer Institut, Villigen, Switzerland}

\date{May 23, 2024}

\begin{abstract}
Neutron imaging has gained significant importance as a material characterisation technique and is particularly useful to visualise hydrogenous materials in objects opaque to other radiations. Particular fields of application include investigations of hydrogen in metals as well as metal corrosion, thanks to the fact that neutrons can penetrate metals better than e.g. X-rays and are at the same time highly sensitive to hydrogen. However at interfaces for example those that are prone to corrosion, refraction effects sometimes obscure the attenuation image, which is used to for hydrogen quantification. Refraction, as a differential phase effect, diverts the neutron beam away from the interface in the image which leads to intensity gain and intensity loss regions, which are superimposed to the attenuation image, thus obscuring the interface region and hindering quantitative analyses of e.g. hydrogen content in the vicinity of the interface or in an oxide layer. For corresponding effects in X-ray imaging, a phase filter approach was developed and is generally based on transport-of-intensity considerations. Here, we compare such an approach, that has been adapted to neutrons, with another simulation-based assessment using the ray-tracing software McStas. The latter appears superior and promising for future extensions which enable fitting forward models via simulations in order to separate phase and attenuation effects and thus pave the way for overcoming quantitative limitations at refracting interfaces.

\end{abstract}

\maketitle

\section{Introduction}
Neutron imaging is a promising analysis tool in many research fields. This is due to the unique penetration capabilities of neutrons and their non-linear dependence of interaction and atomic number \cite{Kardjilov2018}, \cite{Strobl2009_rev}. A key example of this non-linear dependence is the large attenuation cross section hydrogen which renders neutron imaging ideal for observing even small amounts of water or hydrogen-containing compounds in a sample system. Compared to other techniques, such as infrared spectroscopy, secondary ion mass spectroscopy, elastic recoil detection or reflectivity techniques, only neutron imaging allows the determination of hydrogen distribution profile in a micrometer range and is therefore very appropriate to investigate localised corrosion and thick oxide layers on metals, like Al \cite{Ott2020} and Zr \cite{Buitrago2018}.

However, in order to correctly determine the amount of hydrogen in a material the neutron attenuation needs to be measured and analysed accurately. This can be a challenge at the edge or interface of a metallic sample where the neutron refraction takes place due to the corresponding edge enhancement effect, especially in high resolution measurements. In many cases defects or changes in a sample are expected near the interfaces which means the edge enhancement can prevent access to critical information. This differential phase effect depends on a number of factors, like the coherent scattering length of the materials involved, the propagation distance between sample and detector, the neutron energy, i.e. wavelength, the beam collimation, and interface alignment with the beam. Thus, to enable quantitative results based on the neutron attenuation coefficient it is crucial to be able to separate the attenuation signal from the phase effect. This can be done through various phase retrieval procedures developed in X-ray imaging.

Phase retrieval is widespread in X-ray imaging and many different methods have been developed \cite{Langer2008}, \cite{Burvall2011}. Single distance methods are the simplest to implement and useful for tomographies as it is inconvenient and time consuming to measure multiple sample-detector distance for each projection \cite{Burvall2011}. Multiple distance methods, so called holographic methods, are good for quantitative phase contrast, because more information is available to perform the phase retrieval \cite{Cloetens1999}, \cite{Zabler2005}. Another type of holographic method uses multiple wavelength, i.e. energies, instead of propagation distances \cite{Gureyev2001}, \cite{Kashyap2010}.

However, there are key differences between X-ray and neutron imaging, that makes it difficult to use the same phase retrieval methods. The main reason is a lack of coherence of neutron beams due to their relatively low phase space density. In most synchrotron experiments one can assume a highly collimated and coherent beam producing high resolution images at multiple sample-detector distances, whereas with neutrons this is not the case. Images are characterised by increasing blurring with increasing sample-detector distances. Another difference is that for X-rays the necessary propagation-distance to see edge effects is significantly larger than zero, whereas for neutrons the edge effects can be seen even at very small sample-detector distances \cite{Strobl2009}, \cite{Lehmann2017}. Furthermore, the goal for neutron imaging data analysis is rather to retrieve the attenuation signal rather than the pure phase signal, in other words phase contrast filtering rather than phase retrieval. These differences imply that different, novel methods need to be developed to perform phase filtering for neutron imaging. 

In this article we present and compare two phase filtering approaches for attenuation contrast neutron images contaminated with neutron propagation-based phase contrast. To this end, we take data on two reference Al samples, namely a bare and an anodic oxide coated pure Al substrate. Al was chosen for its low neutron attenuation and relatively high coherent scattering, ensuring a high differential phase contrast signal. The choice of a single material and its oxide layer with a simple known geometry aimed to facilitate straightforward modelling of the experiments. The data are measured in two series, at different sample-detector distances and with different wavelengths. The two phase filtering methods considered here are a transport-of-intensity-based phase contrast filter and simulation-based phase filtering through a forward model utilising the neutron ray-tracing software McStas.

\section{Experiments}
\subsection{Samples}
\begin{figure}[h]
    \centering
    \includegraphics[width=0.4\textwidth]{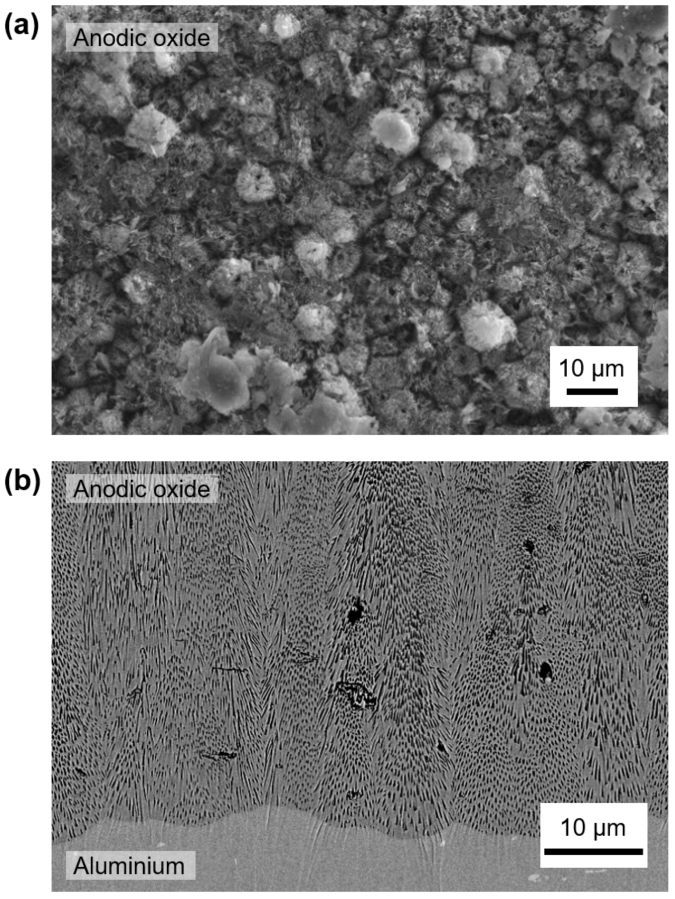}
    \caption{SEM image of anodic oxide coated sample showing (a) the top of the anodic oxide layer and (b) the interface between the anodic oxide and the Al substrate.}
    \label{Fig. SEM}
\end{figure}

Two rectangular samples were imaged with a depth of 2 mm in the beam direction, width of 1 mm and height >25 mm. One consists of pure Al (Al 99.9\%) and the other was an anodised pure Al substrate (Al 99.9\%) with a $225\pm20$\SI{}{\micro\meter} thick layer of Al oxide (\ce{Al2O3}). Both substrates were electropolished at 8 V for 5 min in \ce{HClO4}/ethanol (1/3 v/v) solution at 5\degree C. The samples were rinsed with MilliQ water (Merck \SI{18.2}{\mega\ohm\cm}) and dried with Ar. 
The oxide layer was then grown by double-sided galvanostatic anodising at a current density of 35 mA/cm$^2$ for 9500 s in stirred 0.1 M phosphoric acid at 15\degree C. After anodising, the samples were rinsed with MilliQ water and dried with Ar. For both electropolishing and anodising a two-electrode electrochemical cell was used, the sample acting as the working electrode and a Pt ring electrode as the counter electrode. A Keithley 2400 SourceMeter SMU instrument (Tetronix) was used as a power supply. The oxide layer is porous with an average pore size of $250\pm40$ nm. Porous anodic Al oxide grown in a similar way has been identified as a complex arrangement of Al oxides, oxihydroxides, and hydroxides, referred to as \ce{Al2O3} $\cdot$ n \ce{H2O} \cite{Ott2020} with water content ranging between 5-10\% depending on the distance to the substrate due to variations in porosity. The water includes both water incorporated in the oxide structure and water in the pores. The contribution of the electrolyte anions adsorbed at the anodic Al oxide/ electrolyte interface is neglected because phosphorus possesses rather low neutron cross sections. The average density of the oxide was calculated to be 2.6 g/cm$^3$ and 3.2 g/cm$^3$ near the oxide-substrate interface. SEM images of the resulting oxide layer is displayed in Figure \ref{Fig. SEM}. 
The density variation and inconsistent water distribution induce a significant uncertainty which complicates the forward modelling but also better represents a real sample case.

\subsection{Neutron imaging}
Two separate neutron experiments were performed, one with different sample-detector distances at the Beamline for neutron Optics and other Applications (BOA) at PSI \cite{Morgano2014}, and the other with varying neutron wavelengths at the Imaging with Cold Neutrons beamline (ICON) at PSI \cite{Kaestner2011}. Both the plain and oxide-coated Al samples were measured at each beamline.
The BOA experiment was performed with a white beam with a relevant wavelength range of 0.8-10 Å with a weighted mean wavelength of 3.8 Å. The collimating aperture was square with a side length $D=40$ mm and the aperture-detector distance was $L=7.5$ m, resulting in $L/D = 188$. The measurements were taken at nine sample-detector distances from approximately 1.1 mm to 34.4 mm  with 3.7 mm steps. The sample-detector distances is measured from the centre of the sample. For each distance, 60 radiographs were recorded with exposure time of 30 s each.
The ICON experiment was performed using a velocity selector to select four different wavelengths, specifically 2.7 Å, 3.2 Å, 4.0 Å, and 4.6 Å with a corresponding Gaussian resolution function with a full width at half maximum of $d\lambda/\lambda\approx 15\%$. The sample-detector distance was approximately 1.1 mm. The beam was shaped by a circular aperture with $D=40$ mm diameter and the aperture-detector distance was $L=5.5$ m, which equals a collimation ratio $L/D$ of 138. At each wavelength, 60 radiographs were recorded with an exposure time of 60 s. A summary of the experimental parameters is provided in Table \ref{Tab. exp_param}.

Image acquisition was performed by the PSI Neutron Microscope detector \cite{Trtik2016}, equipped with a 3.5 mm thick isotopically enriched 157-gadolinium oxysulfide scintillator \cite{Crha2019} and a sCMOS camera (Balor, Andor, UK), resulting in a field of view of $9.8\times 9.9$ mm$^2$ and a pixel size of \SI{2.4}{\micro\meter}. This provides a by far sufficient resolution to spatially resolve the oxide and Al layers. Sample alignment with the beam was carefully performed to minimise the influence of sample edge misalignment on the differential phase effects. For both experiments, the final image, $T(x,y)$, was produced by averaging 60 sample images, $I(x,y)$, and normalised by open beam images, $I_O(x,y)$, after subtracting dark current background image $I_D(x,y)$, according to Eq. \ref{Eq. norm}

\begin{gather}
    T(x,y) = \frac{I(x,y) - I_D(x,y)}{I_O(x,y) - I_D(x,y)} \label{Eq. norm}.
\end{gather}

\begin{table}[H]
    \caption{Experimental parameters for neutron imaging experiments.}
    \label{Tab. exp_param}
    \begin{ruledtabular}
        \begin{tabular}{l c c }
                                         & BOA                    & ICON                  \\ 
        \colrule
        Pixel size [\SI{}{\micro\meter}] & 2.4                    & 2.4                   \\
        Aperture                       & 40$\times$40 mm$^2$      & Ø=40 mm                 \\
        Wavelength [Å]                 & 0.8-10                   & 2.7, 3.2, 4.0, 4.6    \\
        Sample-detector distance [mm]  & 1.1 - 34.4               & 1.1   
        \end{tabular}    
    \end{ruledtabular}
\end{table}

\subsection{Phase retrieval and filtering}
Phase retrieval methods are different for the optical near-field and far-field regimes. These regimes are determined by the ratios of the sample width $a$, sample-detector distance $\Delta$, and the wavelength $\lambda$. The near-field regime is defined by $\Delta \ll \frac{a^2}{\lambda}$ and the far-field regime is $\Delta \gg \frac{a^2}{\lambda}$ \cite{Nielsen2011}.

Near field phase retrieval methods are derived by solving a Fresnel integral \cite{Guigay1977}. There are multiple methods of phase retrieval commonly split into two groups, one group is based on the contrast transfer function (CTF) \cite{Cloetens1999} and the other on the transport-of-intensity equation (TIE) \cite{Paganin1998}. The key assumption for CTF phase retrieval is that the sample is a weakly attenuating object inducing a slowly varying phase \cite{Langer2008}. This phase retrieval method needs projections from multiple distances, that are chosen so that the projections map the phase distribution in the Fourier domain. This means choosing distances that span the entire near-field region, so a different number of phase fringes are visible at each distance. This requires a beam coherent enough for multiple phase fringes to be visible and choosing the distances carefully so the projections contain a range of spatial frequencies as represented by the phase fringes. This method also makes quantitative phase retrieval possible due to the sufficient amount of information collected in such an approach.  

The other group of phase retrieval methods are based on TIE considerations, assuming that the propagation distance is short compared to the propagation necessary for CTF filtering \cite{Langer2008}. These methods typically require only one near-field projection. This makes the approach well suited also for images recorded with a less coherent beams but since only one projection is used, only limited phase information might be extracted \cite{Cloetens1999}. On the other hand TIE-based filters are valuable to suppress phase effects and thus enable enhanced attenuation contrast analyses, as is the aim of this work. However, different TIE-based filters require different assumptions about the sample, e.g. that it is a pure phase object or that it is homogeneous and consists of a single material. In addition such filters typically induce additional blur and might obscure small details that are otherwise detectable in the bulk.

Mixed phase retrieval approaches also exist, which combine the CTF and TIE methods. These are based on the assumption of weakly varying objects and also require data from multiple distances resolving multiple phase fringes \cite{Guigay2007}. 

Based on the discussed limitation a TIE-based phase filter appears the best solution amongst these for neutron imaging, in particular the lower coherence requirements match with the available experimental conditions. The specific phase filter used is referred to as the Paganin filter \cite{Paganin2023}, which is widely used in X-ray imaging but is based on the assumption of a homogeneous single-material sample. The goal of the filter is to retrieve images representing the number density of atoms $\rho(\mathbf{r}))$ based on a single material phase and based on the equation 

\begin{gather}
    \rho(\mathbf{r}) = -\frac{1}{\sigma} \log \left( \mathcal{F}^{-1} \left[ \frac{\mathcal{F}[I(\mathbf{r},\Delta)/I_0]}{1+\tau(k_x^2+k_y^2)} \right] \right) . \label{Eq. pag} 
\end{gather}
Here $\mathcal{F}, \mathcal{F}^{-1}$ denotes the Fourier and inverse Fourier transforms, $I(r,\Delta) / I_0$ is the open beam-normalised intensity, $\sigma$ is the total neutron cross section of the material phase, ($k_x$, $k_y$) are the Fourier-space frequencies corresponding to (x,y), and $\tau$ with regards to the specific material is 

\begin{gather}
    \tau = \frac{\lambda^2 b_{\text{coh}}\Delta}{2\pi\sigma} - \frac{(\Theta\Delta)^2}{8} ,
\end{gather}
where $\lambda$ is the neutron wavelength, $b_{coh}$ is the bound coherent scattering length of the materials, $\Delta$ is the sample-detector distance, and $\Theta$ is the beam divergence.

This filter has been derived for neutron imaging, in analogy to the corresponding original X-ray imaging filter \cite{Paganin1998}. The difference between the two lies in particular in the definition of the variable $\tau$ which contains specific relevant material parameters of the sample and the wavelength of the beam in the case of both filters, but only the neutron filter additionally contains the beam divergence. 

This adapted filter has previously been used with neutrons in an attempt to decrease artefacts due to noisy data and increase contrast for monochromatic \cite{Paganin2023} and polychromatic neutron imaging \cite{Oestergaard2023} for cases of pure phase and mixed attenuation and phase contrast neutron imaging data. 

\subsection{Forward model}
We use forward modelling to introduce an additional path towards phase filtering. This approach is based on simulating data as accurately as possible, based on a priori knowledge and the measured data itself and then adapting the sample in the simulation to best possibly fit the measured data. The results enable clear splitting of the phase and attenuation components of the signal. While this is done manually here, we expect successful future implementations to apply fitting routines to this process.

Previously it has been shown that the ray-tracing simulation package McStas \cite{Willendrup2020} qualitatively reproduces experimental data well \cite{Lehmann2017}, \cite{Naver2024}. We chose to model the instrument conditions with McStas with the samples implemented through McStas Union \cite{Bertelsen2017} to simulate the experiments. We used the BOA instrument\cite{BOA_github} which has been benchmarked against measurements \cite{SINE2020} and set the slit configuration and detector to match our experiment. There is no benchmarked ICON instrument, so we implemented a simplified version with a uniform source with the correct divergence and wavelength. 

Differential phase contrast is implemented by refraction at the interfaces described by Snell's law
\begin{equation}
    n_1 \sin\theta_1 = n_2 \sin\theta_2 ,
\end{equation}
where $n_1, n_2$ are the refractive indices of the involved materials at an interface and $\theta_1, \theta_2$ are the angles of the beam with the interface normal. 

In McStas, the neutron refractive index is defined \cite{Sears1982}
\begin{gather}
    n = \sqrt{1 - \frac{\lambda^2\rho b_{\text{coh}}}{\pi}} ,
\end{gather}
where $\rho$ is the number density of atoms.

The two samples were simulated as one and several blocks, respectively. The Al sample was a rectangular cuboid with material parameters given in Table \ref{Tab. sim_param} and of the same dimensions as the real sample. 
The anodic oxide-coated sample was simulated as five layers, namely a pure Al layer in the middle, representing the bulk substrate material, with two oxide layers on either side. The anodic oxide layer was split into two layers in the simulation, in order to account for the non-uniform distribution of porosity and water: oxide layer 1 closest to the Al is modelled as \SI{160}{\micro\meter} wide filled with 10\% water, and oxide layer 2 near the air/oxide interface is considered \SI{80}{\micro\meter} thick with 5\% water. The material parameters used for these layers are also given in Table \ref{Tab. sim_param}. The porosity is reflected in the simulated sample by the oxide material density $\rho_{\text{macro}}$ which is lower than that of the bulk pure Al due to porosity. In addition the Al sample was simulated with a residual rotation angle of 0.24\degree \ in the BOA experiment, to match the asymmetry of the experimental data.

\begin{table}[H]
    \caption{Table of material parameters used for absorption, incoherent scattering, and phase contrast in the wave- and McStas simulations. Neutron cross section and scattering length values for individual elements are from \cite{NIST_neutron}. The values for the oxide layers are calculated as weighted averages between Al and water where the water content is an estimated value based on \cite{Ott2020}.}
    \label{Tab. sim_param}
    \begin{ruledtabular}
        \begin{tabular}{l c c c}
        & Aluminium & Oxide 1 & Oxide 2 \\
        \colrule
        $\mu_{\text{abs}}$ [m$^{^-1}$]   & 1.39   & 0.19  & 0.16   \\
         $\sigma_{\text{inco}}$ [barn]    & 0.0082 & 5.35  & 2.68   \\
         $\sigma_{\text{coh}}$ [barn]     & 1.495  & 2.845 & 3.0225 \\
         $\rho_{\text{macro}}$ [g/cm$^3$] & 2.7    & 2.6   & 2.6
        \end{tabular}    
    \end{ruledtabular}
\end{table}

\section{Results}
\subsection{Aluminium sample}
\begin{figure}[h]
    \centering
    \includegraphics[width=0.4\textwidth]{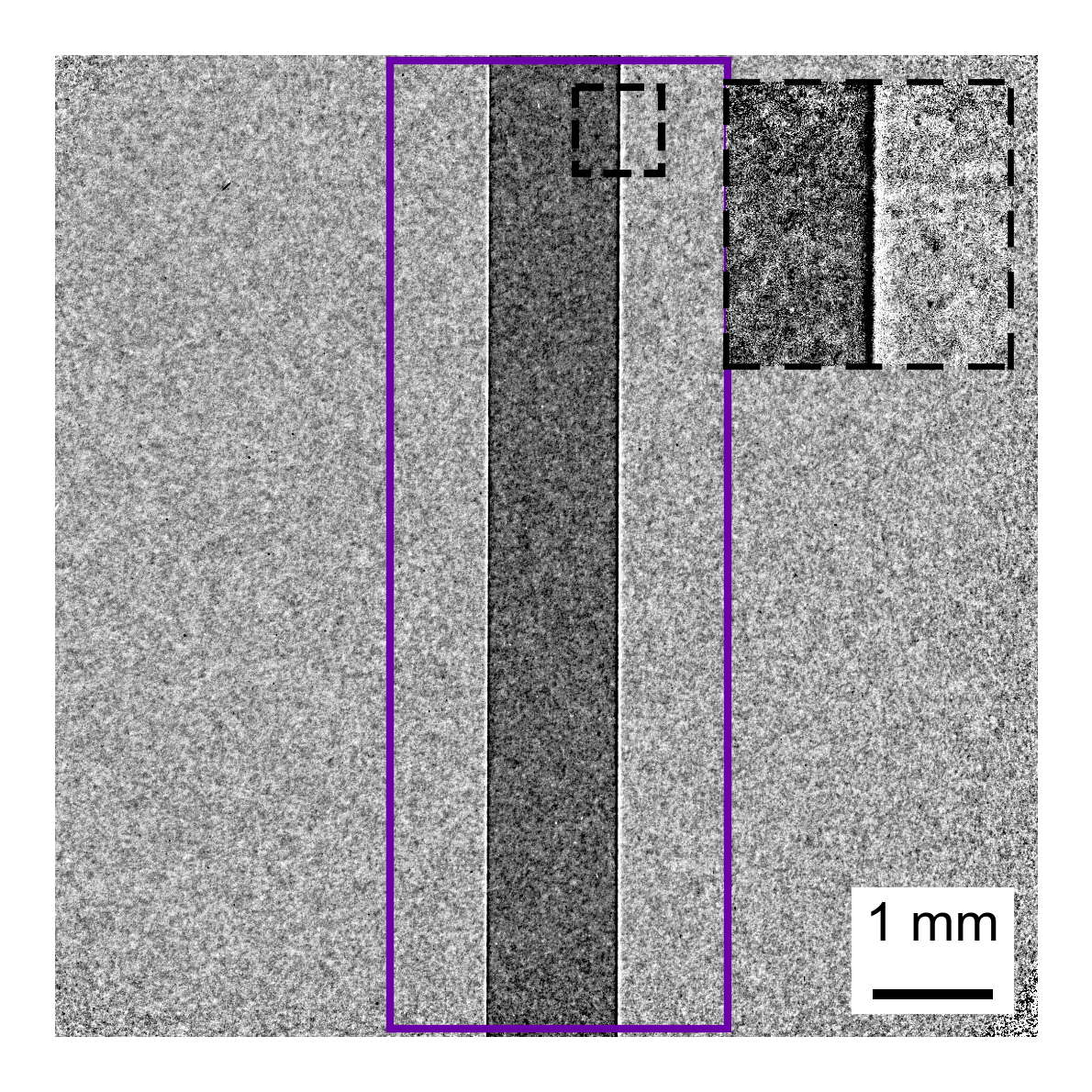}
    \caption{Neutron radiography of Al sample measured at BOA in white beam at 1.1 mm sample-detector distance with a rectangle showing the region of interest. Inset shows zoom of sample edge area in dashed black square.}
    \label{Fig. rad_Al}
\end{figure}

\begin{figure}[h]
    \centering
    \includegraphics[width=0.48\textwidth]{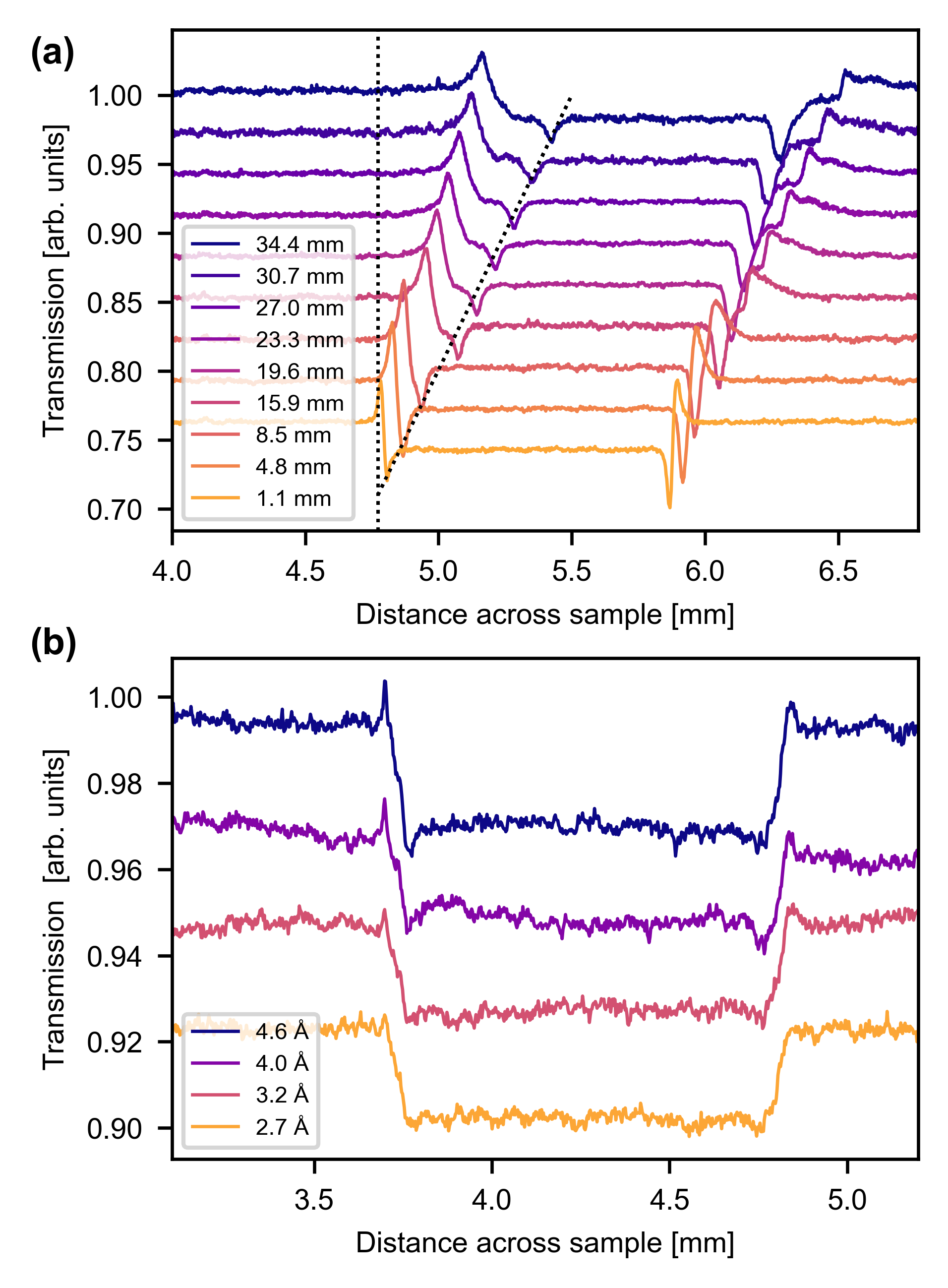}
    \caption{Neutron transmission profile of Al sample at (a) different sample-detector distances in white beam and (b) different wavelengths at $\Delta=1.1$ mm. Dotted line marks the movement of the loss peak. Graphs are offset vertically.}
    \label{Fig. Al_dist_wave}
\end{figure}

A transmission image of the Al sample at $\Delta=1.1$ mm is shown in Figure \ref{Fig. rad_Al} with a rectangle showing the region of interest (ROI). The image was averaged along the sample height in the ROI to calculate the neutron transmission profile, and the resulting profile was plotted for different sample-detector distances and different wavelengths in Figures \ref{Fig. Al_dist_wave}(a) and (b), respectively. The transmission value in air is 1.00 for all samples, but in order to allow better readability, profiles are shown with an artificial vertical offset.

Figure \ref{Fig. Al_dist_wave}(a) shows a clear phase contrast signal at the edges of the Al sample. The signature of the phase effect is a pair of intensity gain and intensity loss peaks. The size and position of these peaks are dependent on both sample-detector distance and neutron wavelength. Both the gain and loss peaks increase in height until $\Delta= 8.5$ mm, after this they widen and flatten out. The distance between the gain and loss peaks increases with increasing sample-detector distance as marked by the dotted line, which together with the asymmetric profile between these peaks suggests that the sample surfaces are not perfectly aligned but at a small angle with respect to the beam direction. The figure also shows a consistent shift of profiles to the right with increased sample-detector distance. This indicates a small misalignment of the sample stage with respect to the beam direction, since this is consistent for all of the distance-dependent BOA measurements. 

\begin{figure}[b]
    \centering
    \includegraphics[width=0.48\textwidth]{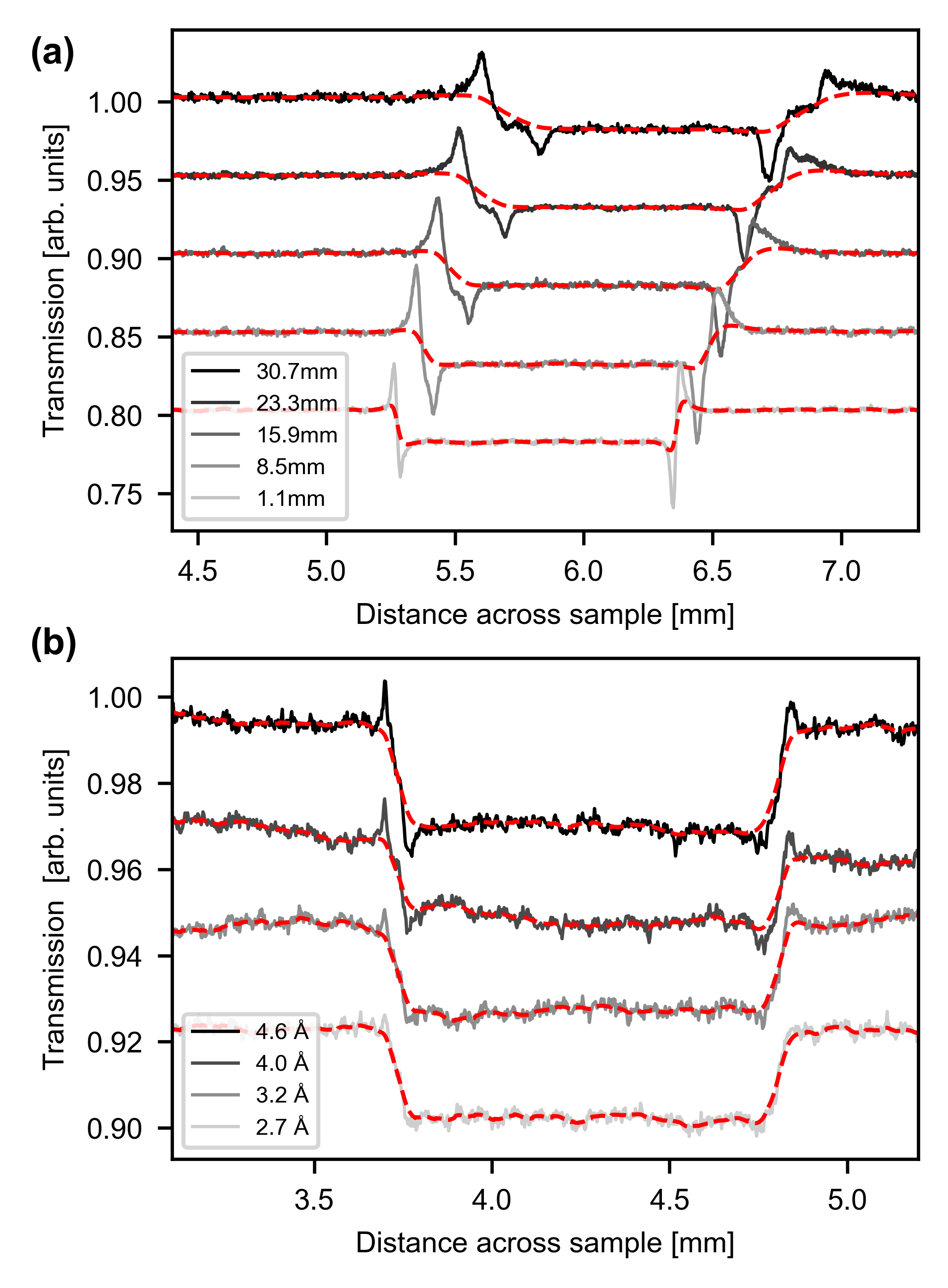}
    \caption{Neutron transmission profile of Al sample at (a) different sample-detector distances in white beam and (b) different wavelengths at $\Delta=1.1$ mm. The solid line plots show experimental data and dashed red shows the data after phase filtering. (a) shows fewer profiles for clarity. Graphs are offset vertically.}
    \label{Fig. Al_paganin}
\end{figure}
Compared to the distance-dependent data, Figure \ref{Fig. Al_dist_wave}(b) shows a significantly weaker phase effect depending on the wavelength. Small and slim loss peaks and larger gain peaks, which both increase in height with increasing wavelength can be observed. It appears that the neutron transmission profile for 2.7 Å contains attenuation contrast only. This is different from the distance-dependent data measured with a broad wavelength spectrum, where even the closest distance results in a pronounced phase peak. 

Subsequently the phase contrast filter defined in Eq. \ref{Eq. pag} was used to perform phase filtering. The results are presented in Figure \ref{Fig. Al_paganin} for the Al sample. The wavelength in Eq. \ref{Eq. pag} was calculated as the weighted mean wavelength of the BOA spectrum for all the BOA measurements. In order to entirely erase the phase effect peaks, the coherent scattering length values used in the filter had to be set 40 times higher than the tabulated values for Al. The figure shows that the profile after phase filtering is more blurred which becomes particularly visible at large sample-detector distances. For clarity, Figure \ref{Fig. Al_paganin}(a) only displays 5 of the 9 recorded profiles, corresponding to every other sample-detector distance measured. The chosen profiles are a representative section of the data, spanning the range of sample-detector distances.

\begin{figure}[b]
    \centering
    \includegraphics[width=0.48\textwidth]{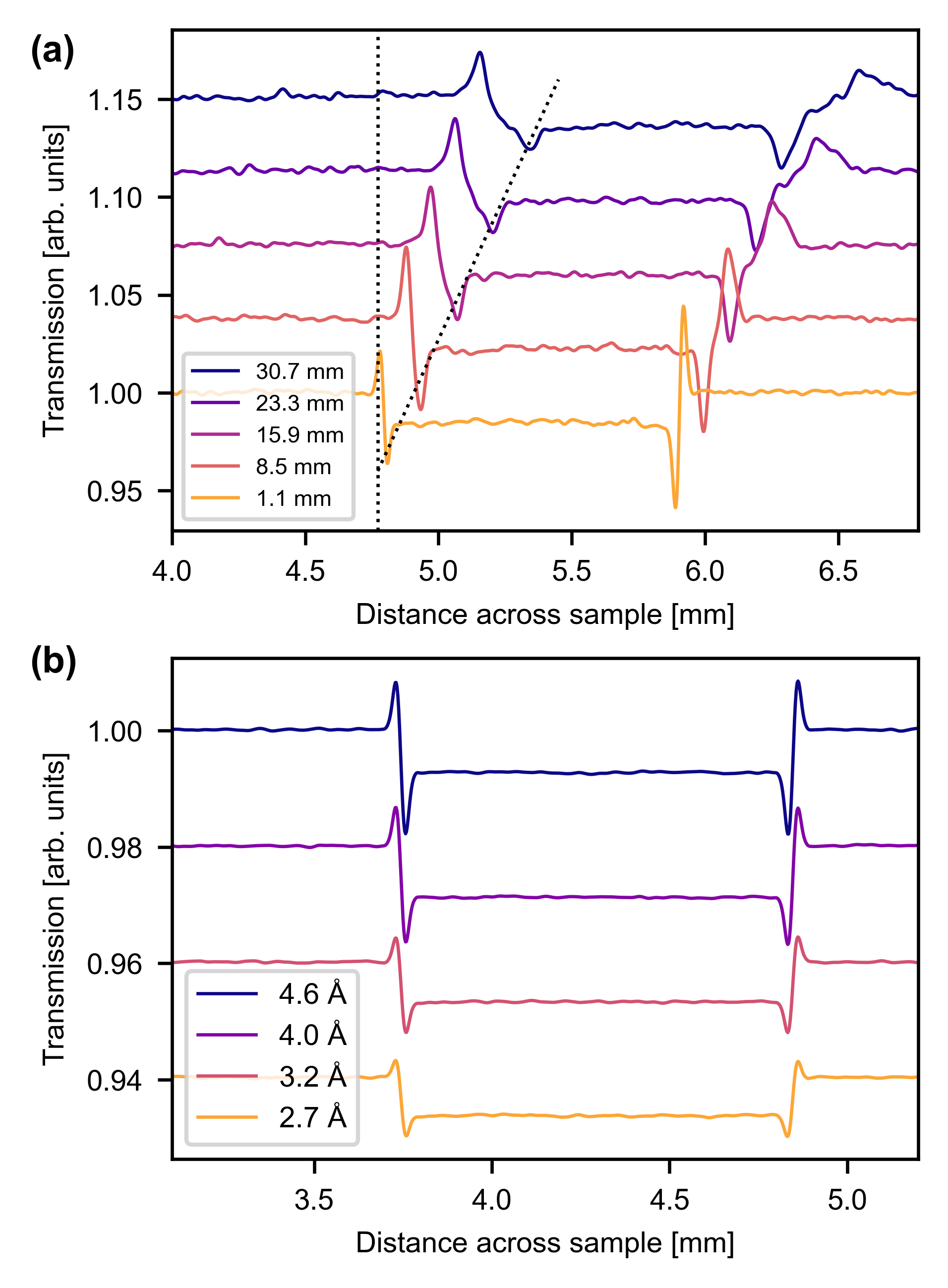}
    \caption{Neutron transmission profile of McStas simulated Al sample at (a) different sample-detector distances in white beam and (b) different wavelengths at $\Delta=1.1$ mm. (a) shows fewer profiles for clarity. Dotted line marks the movement of the loss peak. Graphs are offset vertically.}
    \label{Fig. Al_McStas}
\end{figure}

Similar results are found for the wavelength-dependent data when again using 40x the theoretical coherent scattering length value of Al. The blurring is less obvious as these data were taken at a short sample-detector distance. However, close inspection reveals increasing edge blurring when stronger phase artefacts have to be filtered, i.e. for longer wavelengths, due to the wavelength dependence of the refractive index.

Figure \ref{Fig. Al_McStas}, on the other hand, shows the results of matching the data with simulations. Qualitatively the simulations reproduce the experimental data rather well. The gain and loss peaks move away from each other in the same manner as in the data of the distance scan, as marked by the dotted line, including the asymmetric behaviour, due to the rotational misalignment of the sample, which is accounted for in the simulation. The simulation of the wavelength dependent measurements display a corresponding growth of the phase artefacts with wavelength, however, the 2.7 Å simulation shows a clear signature of phase effects, in contrast to the measurements. This might be a hint of a slightly overestimated collimation and/or resolution in space or with regards to the wavelength resolution.

\begin{figure}[b]
    \centering
    \includegraphics[width=0.48\textwidth]{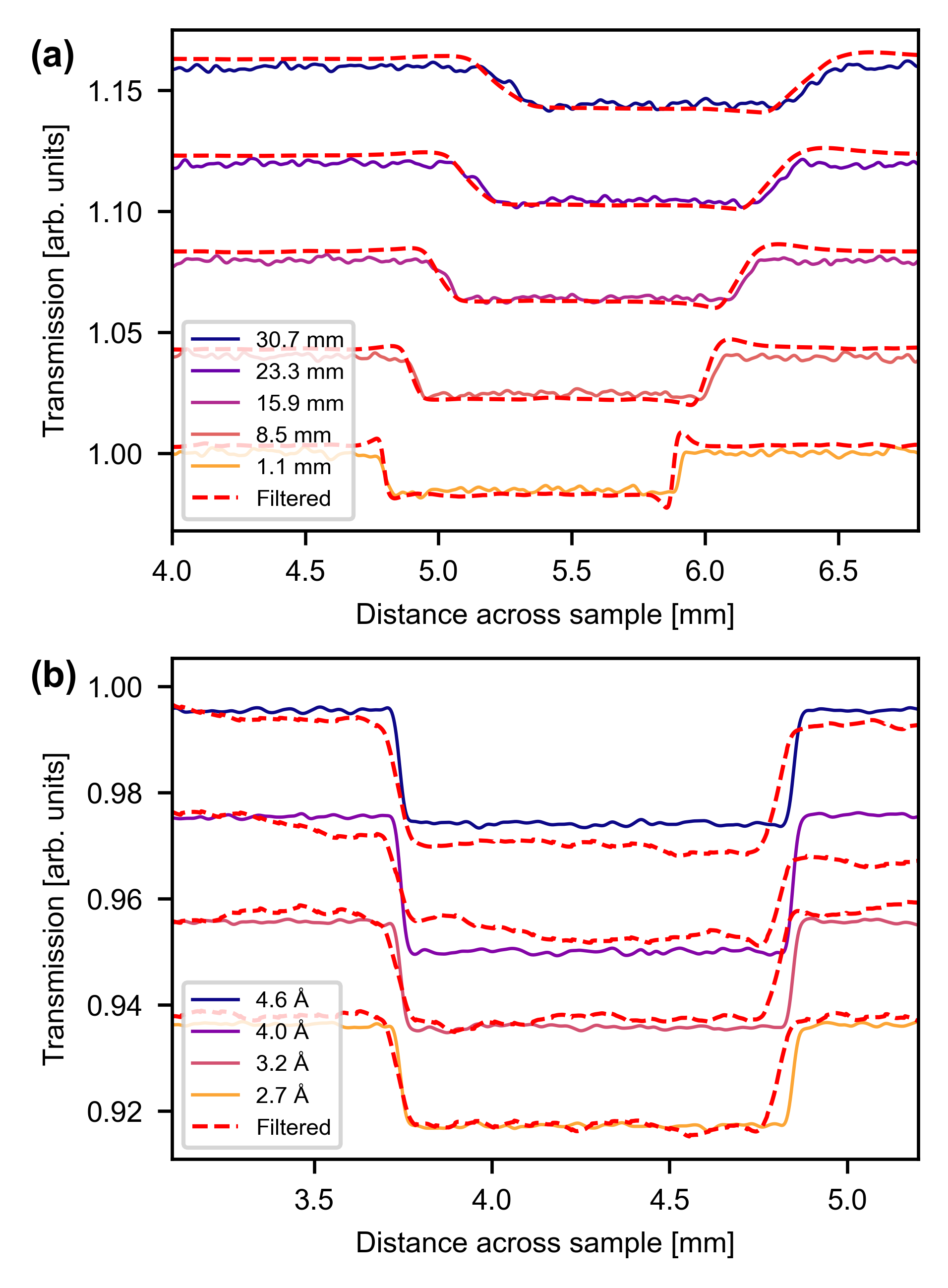}
    \caption{Neutron attenuation profile of Al sample at (a) different sample-detector distances in white beam and (b) different wavelengths at $\Delta=1.1$ mm. The solid lines are McStas simulated attenuation profiles and the red dashed line are experimental data after phase filtering. Graphs are offset vertically.}
    \label{Fig. Al_abs}
\end{figure}

After the applied manual iterating of the McStas simulation model of the sample in order to achieve good agreement between simulation and experimental data, a measurement of the sample simulated without the phase signal was produced. The results are compared to the phase filtered data in Figure \ref{Fig. Al_abs}. The results agree reasonably well with minor contrast variations and apart from some minor phase effect residuals in a few filtered data sets.

\subsection{Anodic oxide coated aluminium sample}
\begin{figure}[h]
    \centering
    \includegraphics[width=0.4\textwidth]{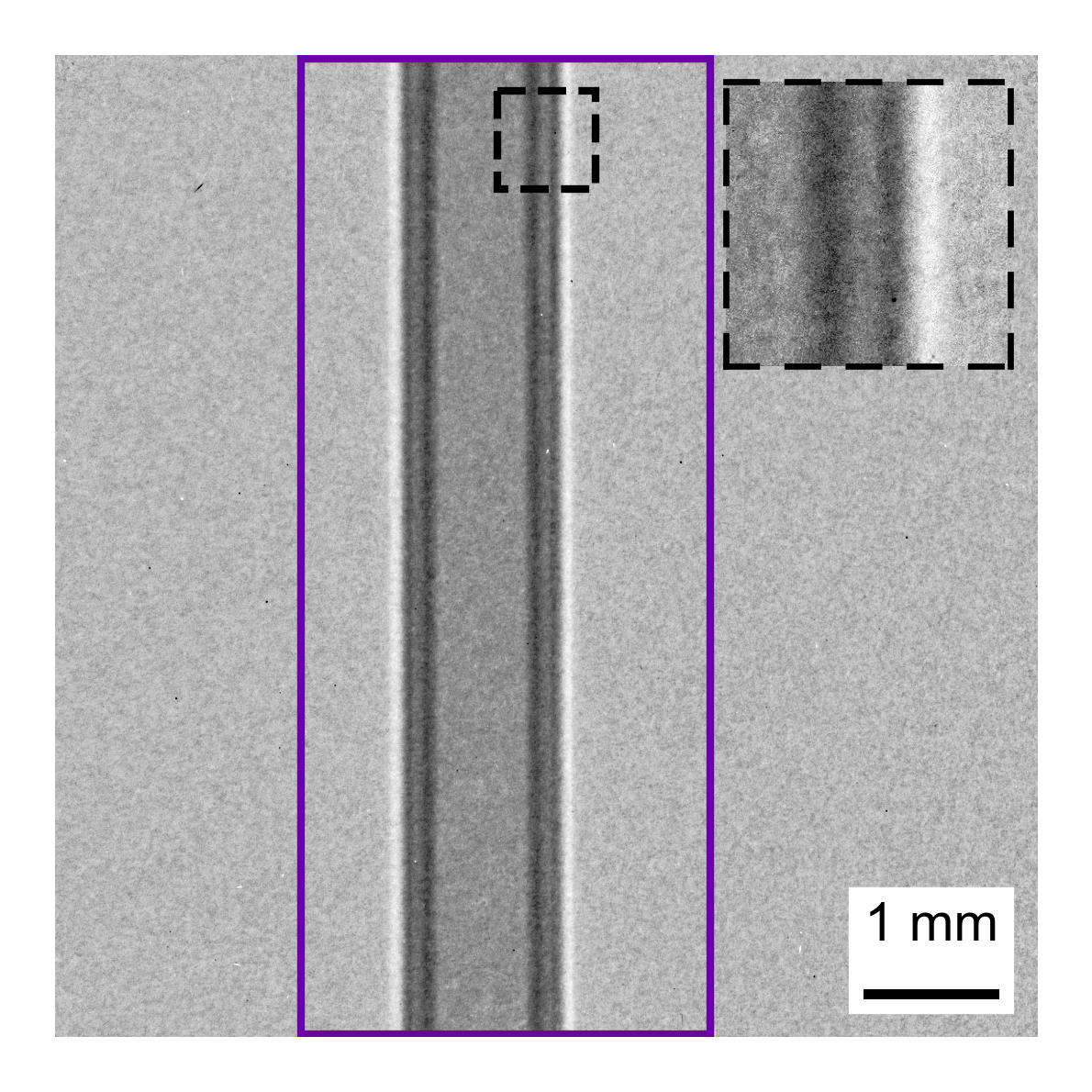}
    \caption{Neutron radiography of Al oxide coated Al sample measured at BOA in white beam at 12.2 mm sample-detector distance. Rectangle showing ROI. Inset shows zoom of sample edge area in dashed black square.}
    \label{Fig. rad_Al_coat}
\end{figure}
\begin{figure}[h]
    \centering
    \includegraphics[width=0.48\textwidth]{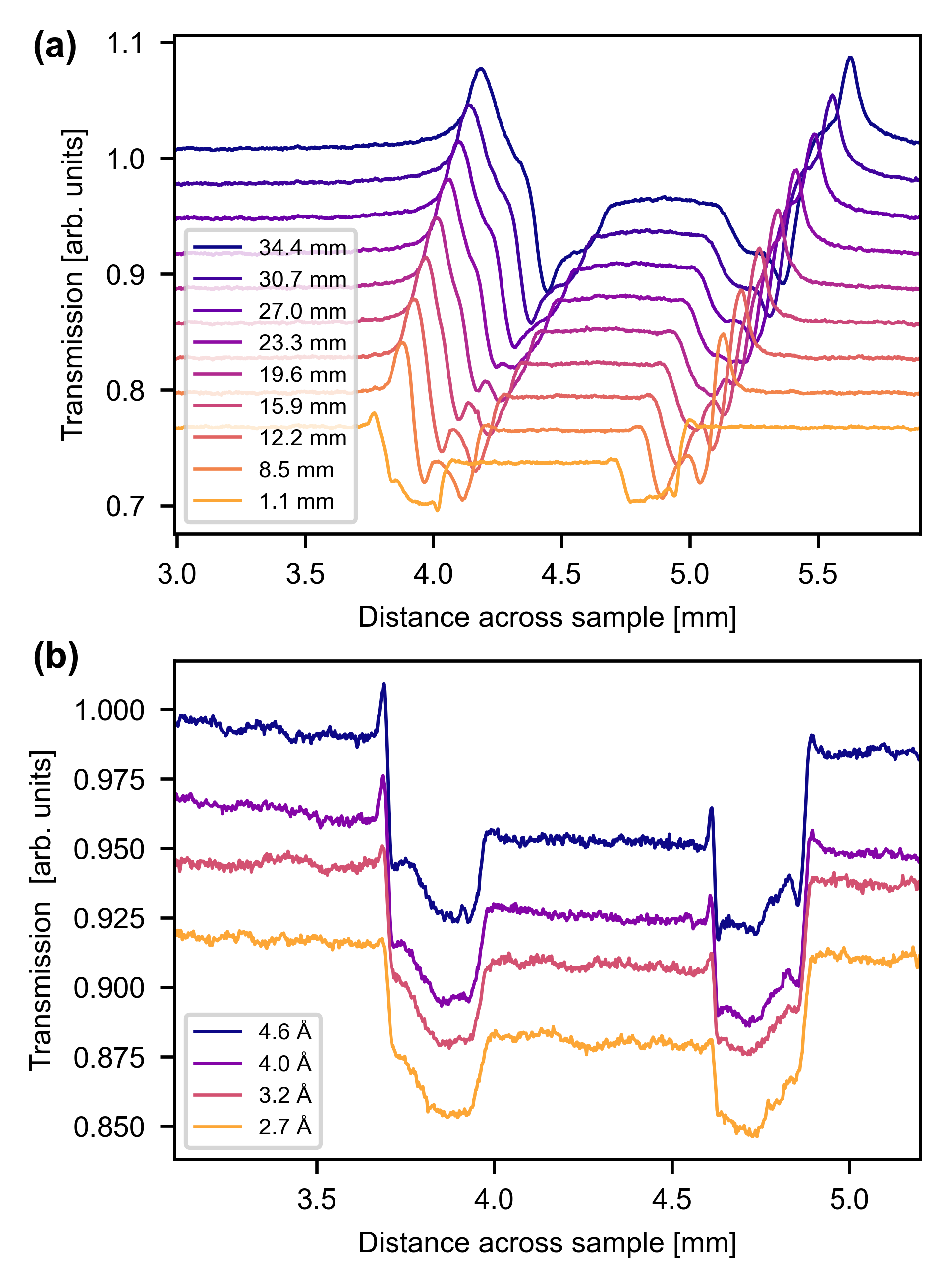}
    \caption{Neutron transmission profile of oxide coated Al sample at (a) different sample-detector distances in white beam and (b) different wavelengths at $\Delta=1.1$ mm. Graphs are offset vertically.}
    \label{Fig. Al_coat_dist_wave}
\end{figure}

A transmission image of the Al oxide coated sample is shown in Figure \ref{Fig. rad_Al_coat} with a rectangle showing the utilised ROI.  The images were again averaged over the height of the ROI to calculate the neutron transmission profiles which are shown in Figure \ref{Fig. Al_coat_dist_wave}.

As already observed in \cite{Ott2020}, the anodic oxide layer has a lower transmission on either side of the bulk Al sample. Figure \ref{Fig. Al_coat_dist_wave} shows that there is a clear gain peak between at the air/oxide interface which increases with sample-detector distance. At the closest sample-detector distance a matching loss peak is not observed but as the distance increases the contrast of the loss peak does too. No phase effect from the oxide/Al interface is observed. At the furthest distance the air/oxide loss peak overlaps more than half with the width of the oxide layer, which makes it difficult to judge how both attenuation and phase contribute to the pattern. A significant fraction of the attenuation difference between the Al and the oxide is biased by the loss peak from the air/oxide interface. 

The data also seems to suggest a non-uniform attenuation signal across the oxide coating. As described earlier the oxide is expected to contain water which is non-uniformly dispersed in the oxide. This appears to be supported by the data. In Figure \ref{Fig. Al_coat_dist_wave}(b) an increase of the air/oxide gain peak with wavelength can be observed as expected, but similar to the $\Delta=1.1$ mm white beam in Figure \ref{Fig. Al_coat_dist_wave}(a) it is difficult to identify the corresponding loss peak. No phase effect is observed at left oxide/Al interface, but narrow gain and loss peaks are seen on the right interface. The peaks are highly asymmetric at the air/oxide and oxide/Al interfaces. 

\begin{figure}[t]
    \centering
    \includegraphics[width=0.48\textwidth]{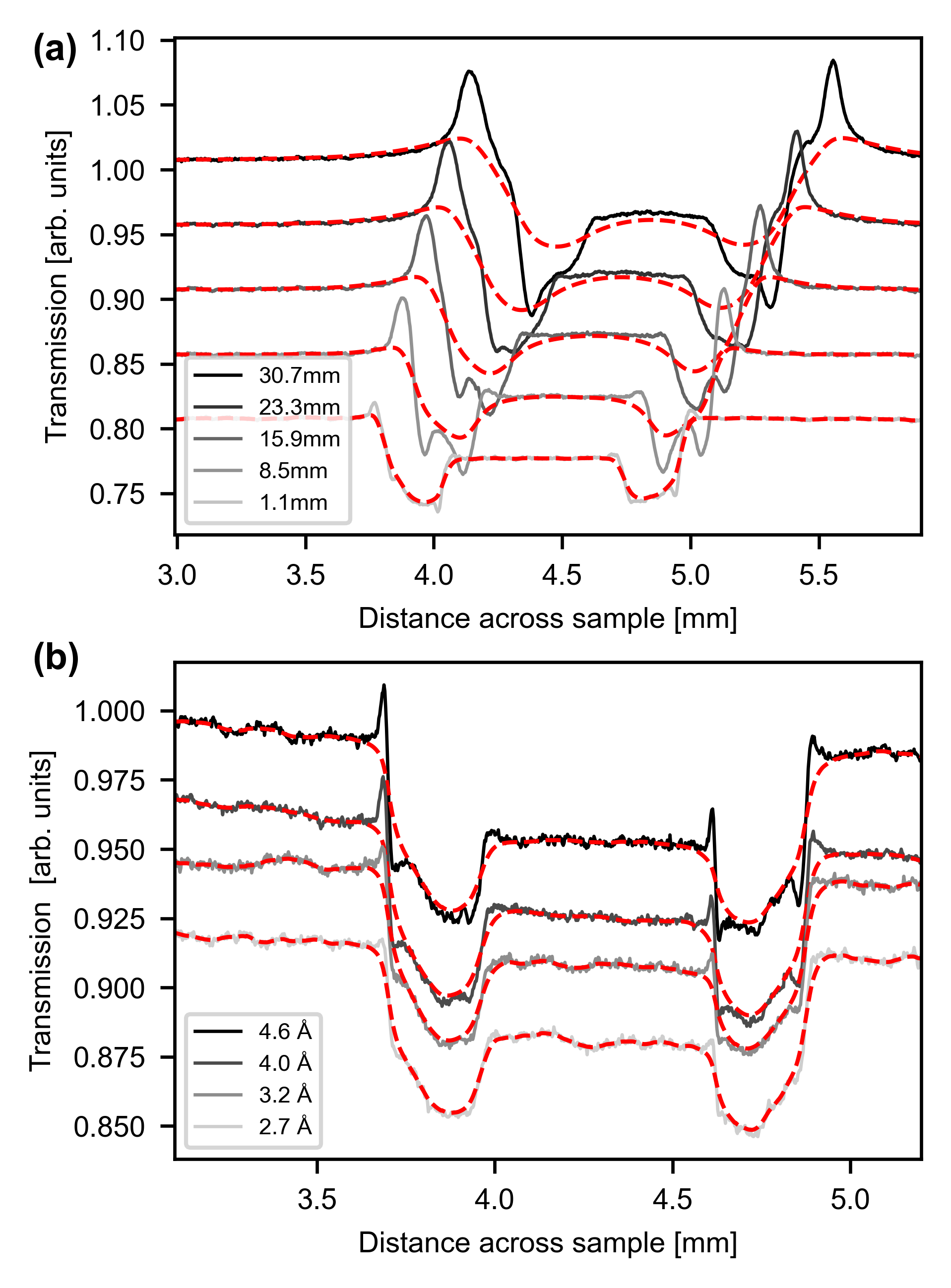}
    \caption{Neutron transmission profile of oxide coated Al sample at (a) different sample-detector distances in white beam and (b) different wavelengths at $\Delta=1.1$ mm. The solid line plots show experimental data and dashed red shows the data after phase filtering. (a) shows fewer profiles, for clarity. Graphs are offset vertically.}
    \label{Fig. Al_coat_paganin}
\end{figure}

Again, the phase filter was applied and the results of this procedure for the oxide coated sample are provided in Figure \ref{Fig. Al_coat_paganin}. As the specific filter includes the assumption of a single-material the parameters of pure Al were applied in the corresponding equation.
The figure shows that after application of the phase filter the sample edges are blurred already at short sample-detector distances and the distinction of oxide coating and the Al substrate is increasingly obscured with increasing sample-detector distance. Even at short sample-detector distances and in the wavelength-dependent data, attenuation differences in the oxide layer are obscured by the filter.
Thus, it becomes obvious that the phase filter does not only remove features due to phase artefacts but also such which can clearly be observed the initial images and are thus significant in terms of attenuation contrast.

\begin{figure}[h]
    \centering
    \includegraphics[width=0.48\textwidth]{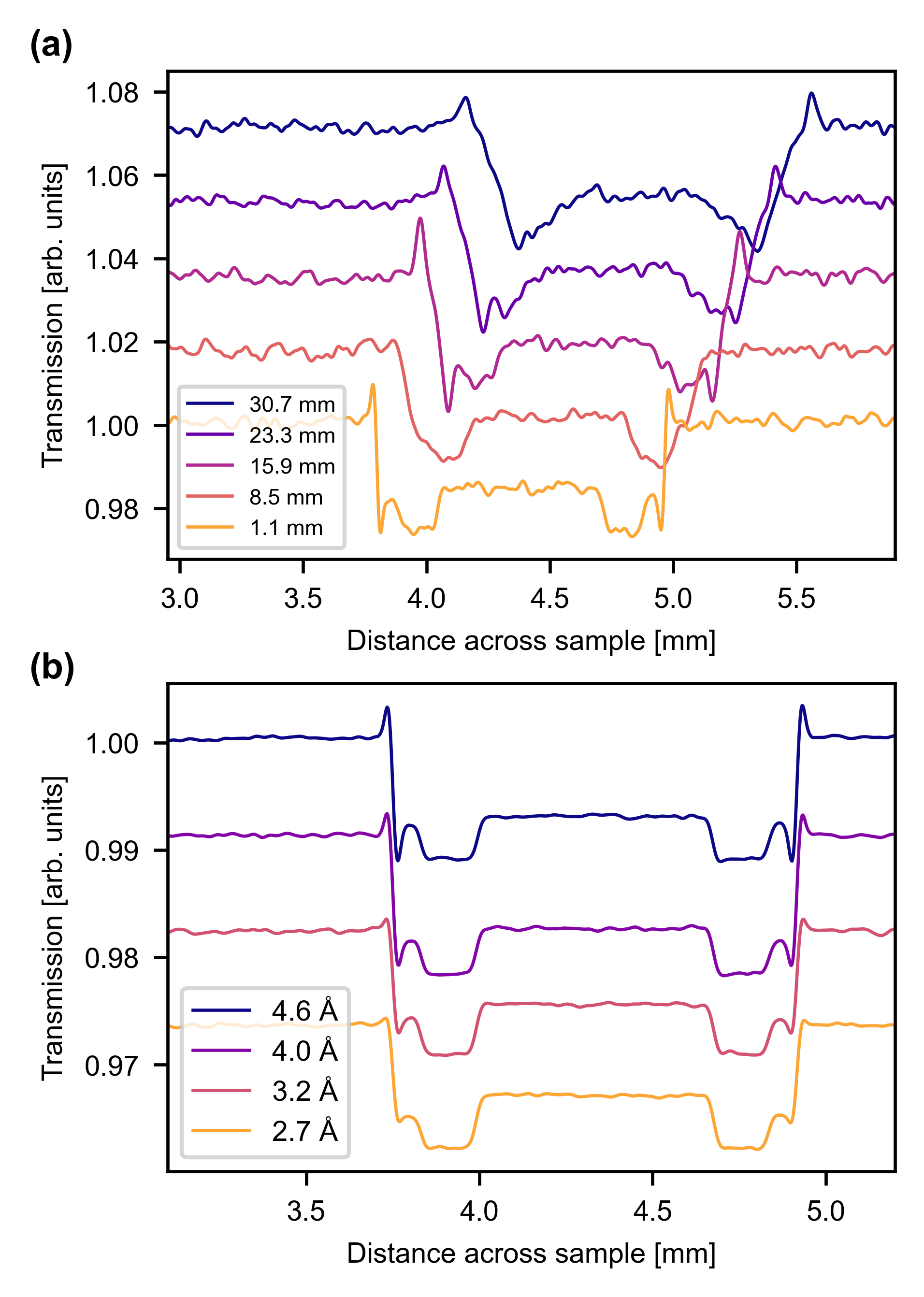}
    \caption{Neutron transmission profile of McStas simulated Al oxide coated Al sample at (a) different sample-detector distances in white beam and (b) different wavelengths at $\Delta=1.1$ mm. (a) shows fewer profiles, for clarity. Graphs are offset vertically.}
    \label{Fig. Al_coat_McStas}
\end{figure}

McStas simulations were run with the five-layer model, described previously, that separate the anodic oxide layer in two different layers. This model is a result of several iterations of the forward model. The simulated profiles shown in Figure \ref{Fig. Al_coat_McStas} reproduce the measured profiles relatively well, in particular with regards to the relative attenuation values. The match between simulation and measurement is best at sample-detector distances above 15.9 mm, though the simulated noise does obscure some features at 30.7 mm. At shorter distances some discrepancies can be observed still.  

The contrast of both air/oxide gain and loss peaks are different in particular at the outer edge of the oxide layer. The loss peak is more distinct in the simulation and gain peak appears more distinct in the measurement of the wavelength-dependent data. It is assumed that sources of deviation are mainly the approximated and simplified structure of the oxide layer concerning density and water content, but potentially also roughness and alignment of the interfaces. 

\begin{figure}[h]
    \centering
    \includegraphics[width=0.48\textwidth]{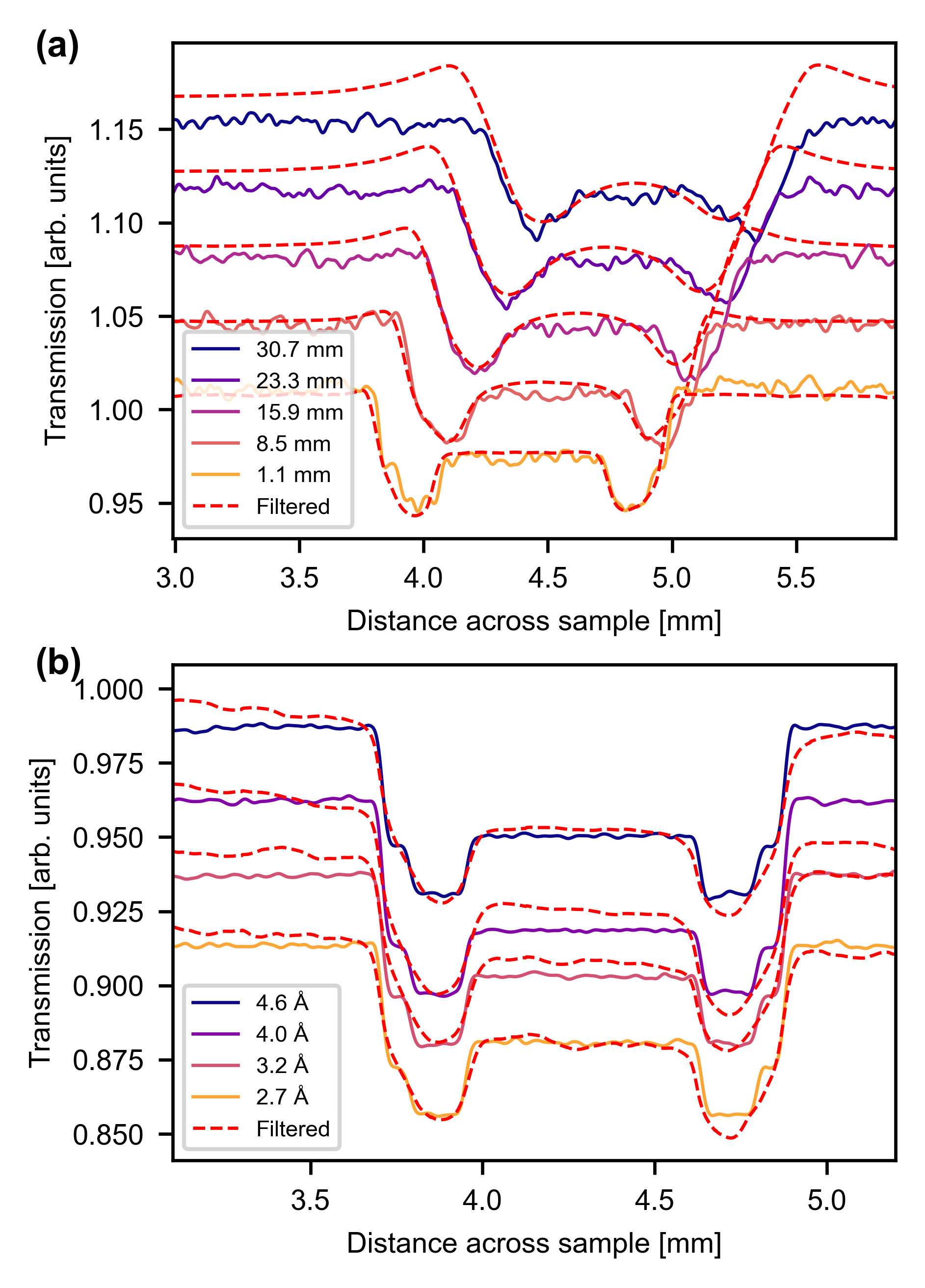}
    \caption{Neutron attenuation profile of Al oxide coated Al sample at (a) different sample-detector distances in white beam and (b) different wavelengths at $\Delta=1.1$ mm. The solid lines are McStas simulated attenuation profiles and the red dashed line are experimental data after phase filtering. The simulated attenuation data has been rescaled to match the phase filtered data. Graphs are offset vertically.}
    \label{Fig. Al_coat_abs}
\end{figure}

Additional simulations of the pure attenuation signal have been performed utilising the five-layer sample model. These are presented together with the phase filtered data in Figure \ref{Fig. Al_coat_abs}. The simulated attenuation profiles have been scaled in order to best possible match the attenuation contrast range of the phase filtered data, in order to enable a qualitative comparison. It can be observed that only the simulated data is able to reveal a layered structure in the oxide layer, for which only a minor hint can be identified in the phase filtered data, specifically the 2.7 Å profile.

\section{Discussion and Conclusion}
The applied neutron phase filter has been shown to work in removing edge effects for a sample of single material phase, namely the simple Al sample. However, it also induces additional blur to the image which is visible at the edges of the sample in high resolution measurements. For a multilayer sample, here the oxide coated Al sample the blurring increases significantly because the algorithm fails to differentiate different material phases and their intertwined contributions to phase effects and attenuation. In the specific case this results in the oxide layer becoming blurred to an extent that it is impossible to distinguish from the bulk Al substrate, instead the profile resembles merely an increased single edge blur. 
For both samples, the filter equation required altered material parameters in order to properly remove the phase artefact peaks. However, it is reported in the X-ray community that it is commonly required to alter the inserted $b$-value in the phase filter, to values differing from the theoretical, tabulated value. This hampers reliability of results even with regards to the desired quantitative attenuation evaluation. The specific phase retrieval filter used is in principle a low-pass filter, which means that high spatial frequencies get filtered and removed from the data. This includes noise just as well as phase contrast artefacts, but also corresponding features of the data representing actual structure in the sample. It is thus very difficult, if possible at all to control the filter through the variables such that the information content is improved rather than diminished.

Another TIE approach \cite{Beltran2010} might be an alternative, in particular concerning dealing with multiple material phases like in the anodic oxide coated Al sample, given correct material parameters for the involved material phases at the respective interfaces can be provided. This might provide better defined interfaces, but it does not solve the other underlying issues of additional blurring and the difficulty of distinguishing subtle attenuation features that should not to be filtered together with the phase artefacts.

The systematic necessity of inputting larger $b$-values for the phase filter in order to remove the phase artefacts, suggests that the filter underestimates the sizes of the phase-induced peaks. While it is unlikely that the actual Al material parameters for differ that much from data base values or that other structural features like porosity and density variations have such an impact, another consideration can be that some of the base assumptions of the filter, developed for X-rays originally, do not hold for neutrons and thus lead to these deviations. Strobl et al. \cite{Strobl2008} conclude in their work about neutron phase contrast artefacts that Fresnel diffraction would lead to smaller edge peaks than refraction- Therefore, the corresponding artefacts in imaging result from refraction and total reflection and not diffraction. Since the phase retrieval filter is built on the description of Fresnel diffraction, the found shortcomings also appear to indicate that Fresnel diffraction causing the edge effect is not a valid assumption for low coherence neutron phase contrast artefacts.

The issues with the phase filtering also imply that X-ray CTF phase retrieval which requires multiple distance measurements and the detection of multiple phase fringes cannot be applied. Again due to coherence limitations multiple phase fringes have never been reported with neutrons and would not support the conclusion concerning mere refraction effects with neutrons. A multiple wavelength version on the other hand, would be ideal for use at pulsed sources, because through the time-of-flight approach multiple wavelength measurements come without flux penalty and might bring a significant efficiency gain, in case wavelength dispersive measurements provide additional necessary information. It should thus be considered to develop such an approach adapted to the case of neutrons and their low available coherence to provide a viable alternative to the computation-intense simulation approach we explore here.

However, here we introduce an approach for correcting phase contrast artefacts which is based on forward modelling through simulations. To this end, we utilise the ray-tracing simulation software package McStas, which is well suited to simulate phase effects in transmission neutron imaging. Initial simulations were based on a priori knowledge and initial assumptions made about the samples. Where the sample material and structure are well known, the use of tabulated material parameters yields good results, in contrast to the phase filter. For more complex and unknown material features, like the porous Al oxide layer of our second sample, assumptions based on first visual inspection of the measured data and subsequently achieved improvements of fitting between model and measurement data were used in a manual approach to fit simulation and measurement. While this is tedious and no perfect agreement can be produced this way, the approach has the potential to be turned into an autonomous fitting procedure in the future. However, we can show that we can improve our sample model and thus develop a qualitatively improved quantification of the measured structure, by eliminating or at least minimising the phase artefact bias. However, one difficulty we met was that the phase effect measured between the porous anodic oxide layer and Al substrate is so weak that we achieved the best fit within the range we could iterate manually, by neglecting it completely. While this is no problem, in terms of retrieving a good attenuation profile, it raises the question of how this can be rationalised, as the bulk material parameters for the involved phases are not suited to understand this. In addition, it can be assumed that the interface is not smooth but affected by significant cracks and porosity as well as potential misalignment with the beam. However, the effect of interface roughness on the respective phase effect cannot yet be simulated in McStas. This implies the need for further experimental investigations with the aim to add such features to the simulation. 

In summary, we have demonstrated that forward simulation-based sample modelling can provide better quantitative approximations of complex interfaces affected by phase contrast artefacts than directly adapted phase filters from X-ray imaging. While the considered phase filter can work for simple interfaces, when manipulating material parameters beyond reasonable values, as soon as there is a subtle structure at the interface the adapted X-ray filters would rather further obscure these structures, than revealing them. On the other hand, when an interface is simple and contains no additional structure, further analyses and quantification are not relevant in any case. Our forward modelling approach through simulations supports improved structure evaluation on more complex interfaces. However, it will only become a valuable alternative and analysis tool, when automation of the approach of simulation and fitting can be achieved, which remains beyond the scope of this work.

\section{Acknowledgements}
We thank the ESS DMSC for use of their simulation cluster. We acknowledge support from the ESS Lighthouse on Hard Materials in 3D, SOLID, funded by the Danish Agency for Science and Higher Education (grant No. 8144-00002B). This work is based on experiments performed at the Swiss spallation neutron source SINQ, Paul Scherrer Institute, Villigen, Switzerland.

\section{References}
\bibliographystyle{ieeetr} 
\bibliography{Bibliography.bib}

\end{document}